\documentclass[12pt]{article}
\usepackage{amsmath,amsfonts,amssymb,graphicx,epsfig}
\date{\today}
\def\unit{\leavevmode\hbox{\small1\kern-3.6pt\normalsize1}}
\newcommand{\be}{\begin{equation}}
\newcommand{\ee}{\end{equation}}
\newcommand{\bea}{\begin{eqnarray}}
\newcommand{\eea}{\end{eqnarray}}

\def \ben{\begin{enumerate}}
\def \een{\end{enumerate}}
\def \bit{\begin{itemize}}
\def \eit{\end{itemize}}

\def\lsim{\raise0.3ex\hbox{$\;<$\kern-0.75em\raise-1.1ex\hbox{$\sim\;$}}}
\def\gsim{\raise0.3ex\hbox{$\;>$\kern-0.75em\raise-1.1ex\hbox{$\sim\;$}}}

\def \vckm{V_{\mathrm{CKM}}}

\def \GeV{{\mathrm{GeV}}}

\def \Im{{\mathrm{Im}}\,}

\def \diag{{\mathrm{diag}}}

\def\21{$SU(2) \ot U(1)$}
\def\ot{\otimes}

\def\bold#1{\setbox0=\hbox{$#1$}
     \kern-.025em\copy0\kern-\wd0
     \kern.05em\copy0\kern-\wd0
     \kern-.025em\raise.0433em\box0 }
\def\nc#1#2#3{{\it Nuovo Cim.}~{\bf#1} (#2) #3}
\def\npb#1#2#3{{\it Nucl.~Phys.\/}~{\bf B #1} (#2) #3}

\def\plb#1#2#3{{\it Phys.~Lett.\/}~{\bf B #1} (#2) #3}

\def\prd#1#2#3{{\it Phys.~Rev.\/}~{\bf D#1} (#2) #3}
\def\prl#1#2#3{{\it Phys.~Rev.~Lett.\/}~{\bf #1} (#2) #3}

\def\hpph#1{{\tt hep-ph/#1}}

\begin{document}
\pagestyle{empty}
\begin{flushright}
FISIST/25-2002/CFIF
\end{flushright}
\vspace{.3cm}
\begin{center}
{\bf{\Large CP violation in supersymmetry with universal strength of
    Yukawa couplings\footnote{Talk given by A.~Teixeira on the
    10$^{th}$ International Conference on Supersymmetry and
    Unification of Fundamental Interactions (SUSY02), June 17-23, DESY
    Hamburg.} }}\\

\vspace* {.5cm}
{\large 
A.M. Teixeira$^\mathrm{a}$, 
G.C. Branco$^\mathrm{a}$,
M.E. G{\'o}mez$^\mathrm{a}$ and 
S. Khalil$^\mathrm{b,c}$ 
}\\
\vspace* {3mm}
$^\mathrm{a}$ {\it Centro de F\'\i sica das Interac{\c c}{\~o}es 
Fundamentais (CFIF),
Departamento de F{\'\i}sica,  Instituto Superior T{\'e}cnico, Av.
Rovisco Pais,  1049-001 Lisboa, Portugal}\\
\vspace* {1mm}
$^\mathrm{b}$ {\it IPPP, Physics Department, Durham University, DH1 3LE,
Durham, U.~K}\\
\vspace* {1mm}
$^\mathrm{c}$ {\it Ain Shams University, Faculty of Science, 
Cairo, 11566, Egypt}

\vspace*{5mm}
{\bf \large Abstract}
\end{center}
{\small
We analyse the CP problem in the context of a supersymmetric extension
of the standard model with universal strength of Yukawa couplings.
In these models we find a small amount of
CP violation from the usual CKM mechanism and 
therefore a significant contribution
from supersymmetry is required.
The electric dipole moments impose severe constraints on
the parameter space, forcing the trilinear couplings to be factorizable in
matrix form. 
We find that the $LL$ mass insertions give the dominant
gluino contribution to saturate $\varepsilon_K$, while chargino
contributions to $\varepsilon^\prime/\varepsilon$ are compatible with
the experimental results. Due to significant supersymmetric
contributions to $B_d-\bar{B}_d$ mixing, the recent large 
value of $a_{J/\psi K_S}$ can be accommodated.}

\section{\bf \large Introduction}
The understanding of the origin of fermion families and the
observed pattern of fermion masses and mixings, together with the
origin of CP violation, are among the major outstanding problems
in particle physics. 

Most extensions of the standard model (SM) naturally include new
sources of CP violation.
In supersymmetric
(SUSY) extensions of the SM we find additional sources of CP
violation, due to the presence of new CP violating phases.
However, these new phases are severely constrained to be small by experimental
bounds on the electric dipole moments (EDM's)~\cite{phase:edm,savoy}.

Since the question of CP-violation is closely related to the
general flavour problem, one may wonder whether it is possible,
within a supersymmetric extension of the SM, to establish a connection
between the need for small CP violating phases and the observed
pattern of quark masses and mixings.
Such a connection might be possible if one
imposes universality of strength for Yukawa couplings (USY) on 
a supersymmetric extension of the SM.
In Ref.~\cite{bgkt}, we introduce the USY ansatz in the framework of
SUSY, analysing the compatibility of this scenario with the EDM
bounds, and discussing the contributions to $K$ and $B$ system CP
violation observables.

\section{\bf \large Universal strength of Yukawa couplings}\label{usy}
One of the most simple, and yet very attractive suggestions
for the structure of the Yukawa couplings consists in assuming that
they have a universal strength $g_y$, so that the mass matrices 
can be written as
$m_{ij}=g_y v/\sqrt{2} \exp(i \phi_{ij})$, where $i,j$ denote family
indices. In the limit of small phases, the USY matrices can be viewed
as a perturbation of the democratic type matrices~\cite{democratic}.
Within the framework of universal strength of Yukawa couplings, 
the quark Yukawa matrices can be
parametrized as 
$U_{ij} = \lambda_u/3\; \exp [i \Phi^u_{ij}] $ and 
$D_{ij} = \lambda_d/3 \;\exp [i \Phi^d_{ij}]$,
where $\lambda_{u,d}$ are overall real constants, 
and $\Phi^{u,d}$ are pure phase matrices. By performing appropriate
weak-basis transformations the matrices $U$ and $D$ can, without
loss of generality, be put in the form
\begin{equation}\label{usy:usy:ansatz}
U = \frac{\lambda_u}{3} \left(
\begin{array}{ccc}
e^{i {p_u}} & e^{i {r_u}} & 1 \\
e^{i {q_u}} & 1 & e^{i {t_u}} \\
1 & 1 & 1
\end{array} \right) \;;
\quad \quad
D = \frac{\lambda_d}{3} K^\dagger \left(
\begin{array}{ccc}
e^{i {p_d}} & e^{i {r_d}} & 1 \\
e^{i {q_d}} & 1 & e^{i {t_d}} \\
1 & 1 & 1
\end{array} \right) K \;,
\end{equation}
where $K=\diag(1, e^{i \kappa_{1}},e^{i \kappa_{2}})$.
The phases ${(p,q,r,t)}$ affect both the
spectrum and the $\vckm$, while $\kappa_i$ only influence the $\vckm$. 
The quark mass hierarchy forces the modulus of
each ${(p,q,r,t)}$ phase to be small, at most of order
$m_2/m_3$. Also, the fact that 
${(|V_{cb}|^2+|V_{ub}|^2)}$ is small, constrains each one of the
phases $\kappa_1, \kappa_2$ to be at most of order
$|V_{cb}|$~\cite{branco:1990-1997, branco:1995}.

In Ref.~\cite{branco:1990-1997}, it was shown that USY can easily
account for the quark mass spectrum at low energy scales.  
This class of Yukawa couplings can be generated at some
high energy scale through the breaking of a flavour symmetry, as
discussed in~\cite{bgkt}, and it has been also recently argued that 
USY textures may naturally arise within the framework of a theory with 
two ``large extra dimensions''~\cite{hung}.
Furthermore, RG evolution preserves 
the USY texture that is responsible for the non-degenerate quark
spectrum~\cite{bgkt}.

By scanning the parameter space, one can find values of 
${(p,q,r,t,\kappa_1,\kappa_2 )}$, such that the quark spectrum and the
$\vckm$ can be correctly reproduced.
To illustrate some of the features of this ansatz, let us
consider, as an example, the following choice of USY phases:

{\small
\begin{table}[h]
\tabcolsep=4.pt
\hspace*{5mm}
\begin{tabular}{ccccccc} \hline
& ${p}$  & ${r}$&
${q}$& ${t}$
& ${\kappa_1}$& ${\kappa_2}$\\
\hline 
up 
& $5.17 \times 10^{-4}$ & $-7.4 \times 10^{-6}$
 & $-1.43 \times 10^{-2}$  & $1.58 \times 10^{-3}$ & $-$ & $-$ \\
down 
& $2.29 \times 10^{-3}$ & $2.83 \times 10^{-2}$
 & $-9.27 \times 10^{-2}$  & $-0.145$ & $0$ & $1.38 \times 10^{-2}$
\\
\hline 
\end{tabular}
\caption{Choice of USY phases for the structure considered in 
Eq.~(\ref{usy:usy:ansatz}).}
\label{table:usy}
\end{table} 
}


The overall factors have been defined as 
$ \lambda_u/3 = m_{t}/v \sin \beta$ and 
$ \lambda_d/3 = m_{b}/v \cos \beta$.
The Yukawa couplings can be diagonalized as
$S_R^{d(u)}~ Y^{d(u)^T}~ S_L^{d(u)^{\dag}} = Y^{d(u)}_{\mathrm{diag}}$,
where $S_R$ and $S_L$ are unitary matrices. It is worth stressing that 
within the context of the SM, the matrices $S_{L,R}^{u,d}$ do not
have, by themselves, any physical meaning, only the combination 
$S_L^u . {S_L^{d^{\dagger}}} \equiv \vckm$ is physically meaningful. 
However the matrices $S_{L,R}^{u,d}$ do play a significant r\^ole in some
extensions of the SM, as for example the MSSM with non-universal
soft-breaking terms. In the present USY model the $\vckm$ is given by 
\begin{equation}\label{usy:ex:vckm2}
\vckm = \left(
\begin{array}{ccc} 
0.975 -6.88 \times 10^{-4} \;i & 0.221+3.96 \times 10^{-3} \;i
& 0.0027 - 4 \times 10^{-5} \;i \\
-0.220 -9.71 \times 10^{-3} \;i & 0.973+0.061 \;i & 
0.041+1.92 \times 10^{-3} \;i \\
0.0063 + 1.10 \times 10^{-4} \;i & -0.041 -1.75 \times 10^{-3} \;i &
0.999+0.027 \;i
\end{array}\right)\;.
\end{equation}

Regarding CP violation, the above $\vckm$ leads to a strength of
CP breaking (measured by 
$J\equiv \Im (V_{ud} V_{cs} V^*_{us} V^*_{cd})$) 
that is too small to account for the observed value of 
$\varepsilon_K$. This is a generic feature of the USY ansatz, and it 
provides further motivation to embed USY in a larger framework, where 
new sources of CP violation naturally arise, like the MSSM.

\section{\bf \large Supersymmetric USY models and CP violation}\label{susy}
In Ref.~\cite{bgkt}, we considered
the minimal supersymmetric standard model (MSSM),
where a minimal number of superfields is introduced and $R$ parity
is conserved, with the following soft SUSY breaking terms 
\bea\label{susy:gen:vsb}
V_{SB} &=& m_{0\alpha}^2 \phi_{\alpha}^* \phi_{\alpha} + 
\epsilon_{ab} 
(A^u_{ij} Y^u_{ij} H_2^b \tilde{q}_{L_i}^a \tilde{u}^*_{R_j} +
A^d_{ij} Y^d_{ij} H_1^a \tilde{q}_{L_i}^b \tilde{d}^*_{R_j} +
A^l_{ij} Y^l_{ij} H_1^a \tilde{l}_{L_i}^b \tilde{e}^*_{R_j} \nonumber\\
&-& B\mu H_1^a H_2^b + \mathrm{H.c.}) 
- \frac{1}{2} 
(m_3\bar{\tilde{g}} \tilde{g} + 
m_2 \overline{\widetilde{W^a}} \widetilde{W}^a +
m_1 \bar{\tilde{B}} \tilde{B})\;,
\eea 
where the Yukawa couplings are of the USY form, and 
$\phi_{\alpha}$ denotes all the scalar fields of the theory.
The $\mu$ term and the gaugino soft terms are assumed to be real, so
that the structure and phases of the trilinear terms will play a key
r\^ole, regarding both the stringent bounds coming from the EDM's, and
the enhancement of the CP observables.
In this work we will also take as a guideline the assumption that 
all supersymmetric phases should be no greater than the largest of the USY
phases in Table~\ref{table:usy}.

We begin by investigating how the EDM bounds constrain the
parameter space for SUSY models with universal strength of Yukawa 
couplings. The current experimental bound on the EDM of the neutron
and mercury atom are 
$d_n < 6.3 \times 10^{-26}\; e\;\text{cm}$ and
$d_{Hg} < 2.1 \times 10^{-28}\; e\;\text{cm}$, 
respectively~\cite{edm:neutron:Hg}.
These bounds can be translated into constraints for the imaginary parts of the 
flavour conserving $LR$ mass insertions. 
Requiring that $d_n$ does not exceed the experimental limit 
compels $\Im (\delta^{d(u)}_{11})_{LR}$ to be less than $10^{-6}$,
while compatibility with the mercury EDM corresponds to having 
$\mathrm{Im}(\delta^{d(u)}_{11})_{LR} \lsim 10^{-7}~
~-~10^{-8}$ and $\mathrm{Im}(\delta^{d}_{22})_{LR} \lsim 10^{-5}
-10^{-6}$~\cite{phase:edm}. 

\begin{figure}[ht]
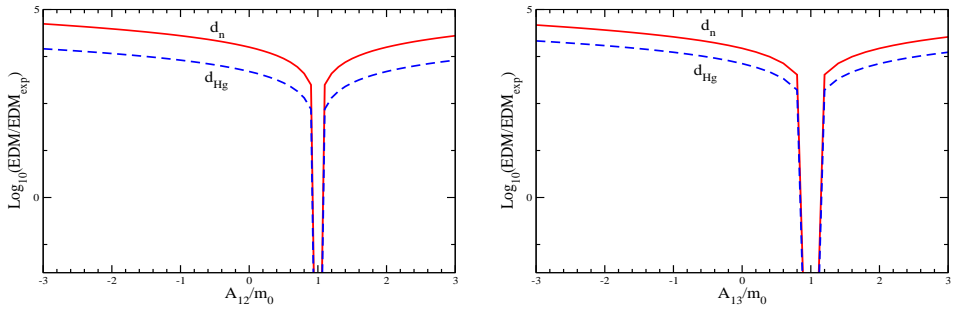

\begin{center}
\hspace*{-7mm}
\epsfig{file=na12.eps,width=6cm,height=4cm, clip=} \quad
\epsfig{file=na13.eps,width=6cm,height=4cm, clip=}\\
\caption{Neutron and mercury EDM's as function of the off-diagonal 
entries $A_{12}$ and
$A_{13}$ for $\tan \beta=5$ and $m_0=m_{1/2}=250$ GeV. 
$A_{ij}=1\;\forall ij \neq 12 (13)$ on the left (right) figure.}
\label{susy:edm:fig1}
\end{center}
\end{figure}

We showed that, even in the limit of small (or vanishing) supersymmetric
phases, the large mixing inherent to USY couplings
(displayed in $S_L$ and $S_R$), together with the EDM experimental
limits, severely constrain any non-universality of the trilinear terms.
In Fig.~\ref{susy:edm:fig1} we present the constraints from the 
EDM's on the off-diagonal entries of the $A$-terms, in particular 
$A_{12}$ and $A_{13}$. In our analysis we assumed $\tan \beta=5$,
$m_0= m_{1/2}= 250\;\GeV$ and 
$A_{ij} = m_0$ for all elements except $A_{12,13}$, which we set in
the range $[-3 m_0, 3 m_0]$.
\vspace*{5mm}

From these figures, it is clear that any significant 
non-universality among the $A$ terms
leads to unacceptably large contributions to the
EDM's. Similar constraints hold for the other 
off-diagonal elements $A_{21}$, $A_{31}$, $A_{23}$ and $A_{32}$, and
this situation is far more severe than the case of hierarchical Yukawa
couplings.
Since Hermitian USY couplings lead to an unrealistic sum rule for the
quark masses, the associated EDM 
problem cannot be overcome by taking simultaneously
Hermitian Yukawa and trilinear couplings~\cite{real:yukawa}.
An interesting possibility is to have trilinear
terms that can be factorized as $\hat{A} = A_{ij} Y_{ij} = A.Y$ or 
$Y.A$~\cite{vives}. This factorization implies that the 
mass insertion $(\delta_{11}^d)_{LR}$ is 
suppressed by the ratio $m_d/m_{\tilde{q}}$. To be specific, 
let us consider the following structure
\begin{equation}\label{Aterms}
A = m_0 \left(\begin{array}{ccc}
a & a & a \\
b & b & b \\
c & c & c \end{array} \right) \;.
\end{equation}
In this case the trilinear couplings $\hat{A}$ can be written as 
$\hat{A} = \mathrm{diag}(a,b,c) . Y$, with complex 
$(a,b,c)$. 
For $\vert A_{ij} \vert \simeq 3 m_0$, the EDM's constraint
on these phases is quite stringent: 
\begin{equation}\label{susy:edm:dhgm0}
\varphi_a \lsim 0.02 \; \mathrm{rad} \quad \;; \quad \quad 
\varphi_{b} \lsim 0.35 \; \mathrm{rad}\quad \;; \quad \quad 
\varphi_c \lsim 0.02 \; \mathrm{rad}\;.
\end{equation}
Thus, we can conclude that in USY models the maximal allowed
non-universality for the $A$-terms is the structure presented in  
Eq.~(\ref{Aterms}), with the associated SUSY phases
($\varphi_{ij}$) within the limits given in Eq.~(\ref{susy:edm:dhgm0}).
In view of this, we analyse the contributions to the kaon
system CP violating observables, 
$\varepsilon_K$ and $\varepsilon^\prime/\varepsilon$.

Let us start our analysis by considering 
the indirect CP violating parameter of the $K$ sector, 
$\varepsilon_K$.
In the presence of supersymmetric ($\tilde{g}$ and $\tilde{\chi}^\pm$) 
contributions,
the off-diagonal entry in the kaon mass matrix can be decomposed as 
$M_{12} = M_{12}^{\mathrm{SM}} +  
M_{12}^{\tilde{g}} +  M_{12}^{\tilde{\chi}^\pm}$. 
Within the present scenario, 
with USY phases as in Table~\ref{table:usy}, 
we find that the SM contribution to $\varepsilon_K$ is $\mathcal{O}(10^{-5})$. 
Regarding the supersymmetric contributions, and taking the 
$A$-terms as in Eq.~(\ref{Aterms}), we have studied the behaviour of
the $(\delta_{12}^d)_{LL}$, $(\delta_{12}^d)_{LR}$ and 
$(\delta_{12}^u)_{LL}$, mass insertions, focusing on the correlation
between the $(\delta_{12}^d)_{LL}$ and 
$(\delta_{12}^d)_{LR}$. We have verified that gluino contributions
clearly dominate over those of the charginos, and 
that the $LL$ mass insertions provide the dominant 
gluino contribution.
We notice that in 
our model, the large mixing displayed by the rotation
matrices enhances any non-diagonal contribution to $M^2_{\tilde{Q}_L}$
induced by the trilinear terms from RG evolution, hence $LL$
mass insertions can account for the experimental value of $\varepsilon_K$.

We then consider supersymmetric 
contributions to the direct CP violation parameter,
$\varepsilon^\prime/\varepsilon$, which can be symbolically
decomposed as a sum of the SM, gluino, and chargino terms, the
latter receiving contributions from diagrams with one and two mass
insertions in the squark internal line. 

\begin{figure}[ht]
\begin{center}
\epsfig{file=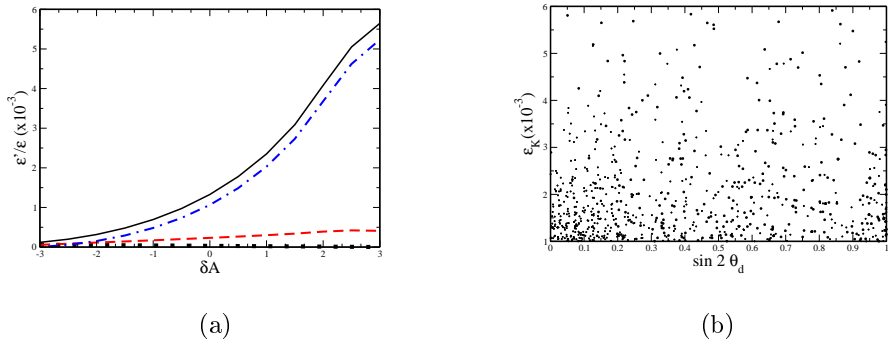}\\
\caption{(a) Contributions to $\varepsilon^\prime/\varepsilon$ versus
the non-universality parameter $\delta A$.
The dotted line refers to gluino contributions, the
dot-dashed and dashed lines to those of the charginos 
(with one and two mass insertions, respectively). The solid
line stands for the total supersymmetric contribution 
to  $\varepsilon^\prime/\varepsilon$.
(b) Correlation between $\varepsilon_K$  and $\sin 2 \theta_d$.}
\label{ep_2b}
\end{center}
\end{figure}

In USY scenarios, the SM prediction for $\varepsilon^\prime/\varepsilon$
is found to be $\mathcal{O}(10^{-6})$.
The supersymmetric contributions were analysed with detail
in~\cite{bgkt}. In order to discuss the r\^ole of each contribution within
our model, in Fig.~\ref{ep_2b}(a)
we plot the gluino and chargino contributions to 
$\varepsilon^\prime/\varepsilon$ as function of $\delta A$, 
which is defined as 
\begin{equation} 
\delta A = (A_{3i} - A_{1i,2i})/m_0 = c-a \quad (\mathrm{assuming}\;a=b) \;.
\end{equation}
The SUSY CP violating phases have been fixed as: 
$\varphi_a=\varphi_c=0$ and $\varphi_b=0.1$. 
As in the previous figures, we have assumed $\tan \beta=5$ and 
$m_0=m_{1/2}=250$ GeV. 
\vspace*{2mm}

As can be seen from this figure, the chargino contributions with one mass 
insertion give the dominant contribution to $\varepsilon^\prime/\varepsilon$.
The chargino contributions with two mass insertions that arise from the 
SUSY effective $\bar{s}dZ$ vertex are also relevant and considerably
larger than that of the gluino.

Finally, we address the issue of CP violation in the $B_d$ meson
system.
The present world average, $a_{J/\psi K_S}=0.79\pm0.12$, is dominated
by the BaBar and Belle results~\cite{babar:belle}. 
In the case of supersymmetric contributions to
the $\Delta B= 2$ transition, the ratio of the total and partial SM
contributions can be parametrized as 
$r_d^2 e^{2 i \theta_d} =
M_{12}(B_d)/M_{12}^{\mathrm{SM}}(B_d)$, hence 
the measurement of $a_{J/\psi K_S}$ does not determine $\sin 2 \beta$ 
but rather $\sin(2 \beta + 2 \theta_d)$.
The SM contributions to $\sin 2 \beta$ associated with the phases
of Table~\ref{table:usy} are negligible, $\mathcal{O}(10^{-3})$.

Regarding SUSY contributions, we took into account charged-Higgs,
chargino and gluino contributions to 
$M_{12}^{\mathrm{SUSY}}(B_d)$.
Since in this class of models flavour mixing is not suppressed, one finds
that the $(\delta^{d,u}_{13})_{LL}$ and $(\delta^{d,u}_{13})_{LR}$ mass
insertions 
can be significantly large and saturate the experimental result. 
We have found that within our scenario, the leading supersymmetric 
contribution to $\sin(2 \beta + 2 \theta_d)$ was associated with the 
$\Delta B=2$ chargino mediated box diagrams.
In Fig.~\ref{ep_2b}(b) we present the correlation between the values of
$\varepsilon_K$ and $\sin 2 \theta_d$ for $\tan \beta =5$ and $m_0=m_{1/2}=
250$~GeV. The values of the parameters $a,b,$ and $c$ are randomly selected
in the range $[-3,3]$ and the phases fixed as $\varphi_{a}=\varphi_{c}=0$
and $\varphi_b=0.1$.
From Fig.~\ref{ep_2b}(b), 
it is clear that one can have $\sin 2 \theta_d$ within
the experimental range 
while having a prediction for $\varepsilon_K$ and
$\varepsilon'/\varepsilon$ compatible with the measured value. 
This result appears as characteristic of the 
kind of models under consideration. Despite the smallness of the phases 
introduced in these models via the Yukawa and trilinear couplings, 
they are sufficient to account for CP
violation in the $K$-system, as well as the observed value of 
$a_{J/\psi K_S}$.

\section{\bf \large Discussion and Conclusions}
In this work we have studied the implications of having universal
strength of Yukawa couplings within the unconstrained MSSM.

We have argued that the trilinear soft terms play a key r\^ole in
embedding USY into SUSY. In fact, due to the large mixing and
associated phases, 
the constraints from the EDM's on the SUSY parameter space are far
more stringent than in the case of a standard Yukawa parametrization.
We found that in order to satisfy the bound of the mercury EDM, the
$A$-terms should be matrix factorizable, with phases constrained
to be in the range $10^{-2}-10^{-1}$.

We have investigated the new
contributions to both $K$ and $B$ system CP observables,
finding that gluino mediated boxes with $LL$ mass
insertions provide the leading contributions to $\varepsilon_K$, while 
$\varepsilon^\prime/\varepsilon$ is dominated by chargino loops,
through $LL$ flavour mixing.
Regarding the $B$ system, we argued that within this model 
supersymmetric 
chargino exchanges,
provide the leading contributions, which are in agreement with  
the recent measurements at BaBar and Belle.

In conclusion, we have presented an alternative scenario for CP
violation, where the 
strength of CP violation stemming from the SM is naturally small,
so that new SUSY contributions are 
essential to generate the correct value of 
$\varepsilon_K$ and  $\varepsilon^\prime/\varepsilon$, as well as 
the recently observed large value of $a_{J/\psi K_S}$.\\

\noindent {\bf \large Acknowledgments}

This work was supported in part by
the Portuguese Ministry of Science through project CERN/P/Fis/40134/2000, 
CERN/P/Fis/43793/2001, and by the E.E. through project HPRN-CT-2000-001499.
M.G. and A.T. acknowledge support from 'Fundac\~ao para a Ci\^encia e
Tecnologia', under grants SFRH/BPD/5711/2001
and PRAXIS XXI BD/11030/97, respectively.  
The work of S.K. was supported by PPARC.


{\small
\begin{thebibliography}{99}
\bibitem{phase:edm} 
S.~Abel, S.~Khalil and O.~Lebedev, 
\npb{606}{2001}{151}.
\bibitem{savoy}
S.~Pokorski, J.~Rosiek and C.~A.~Savoy, \npb{570}{2000}{81}.
\bibitem{bgkt}
G.~C.~Branco, M.~E.~G\'omez, S.~Khalil and A.~M.~Teixeira, 
\hpph{0204136}.
\bibitem{democratic} 
H.~Fritzsch and J.~Plankl, 
\prd{49}{1994}{584};
H.~Fritzsch and P.~Minkowski, 
\nc{30A}{1975}{393};
H.~Fritzsch and D.~Jackson,
\plb{66}{1977}{365};
P.~Kaus and S.~Meshkov,
\prd{42}{1990}{1863}.
\bibitem{branco:1990-1997} 
G.~C.~Branco, J.~I.~Silva-Marcos and M.~N.~Rebelo,
\plb{237}{1990}{446};
G.~C.~Branco, D.~Emmanuel--Costa and J.~I.~Silva-Marcos, 
\prd{56}{1997}{107}.
\bibitem{branco:1995}
G.~C.~Branco and J.~I.~Silva-Marcos, 
\plb{359}{1995}{166}.
\bibitem{hung} 
P.~Q.~Hung and M.~Seco,
\hpph{0111013}.
\bibitem{edm:neutron:Hg}
P.~G.~Harris {\it et al}, 
\prl{82}{1999}{904};
M.~V.~Romalis, W.~C.~Griffith and E.~N.~Fortson, 
\prl{86}{2001}{2505};
J.~P.~Jacobs {\it et al.}
\prl{71}{1993}{3782}.
\bibitem{real:yukawa}
S.~Abel, D.~Bailin, S.~Khalil and O.~Lebedev,
\plb{504}{2001}{241};
S.~Khalil, \hpph{0202204}.
\bibitem{vives}
S.~Khalil, T.~Kobayashi and O.~Vives,
\npb{580}{2000}{275};
T.~Kobayashi and O.~Vives,
\plb{506}{2001}{323}.
\bibitem{babar:belle}
BABAR Collaboration, B. Aubert {\it et al.},
\prl{87}{2001}{091801};
BELLE Collaboration, K. Abe {\it et al.},
\prl{87}{2001}{091802}.

\end{thebibliography}
}
\end{document}